\documentclass[prl,floatfix,twocolumn,showpacs,amsmath,amssymb]{revtex4}

\usepackage{graphicx}
\usepackage{dcolumn}
\usepackage{bm}

\begin{document}
\title{
Magic angle effects of the one-dimensional axis conductivity in quasi-one dimensional conductors
}

\author{Yasumasa Hasegawa$^{1}$,  Hirono Kaneyasu$^{1}$, and  Keita Kishigi$^{2}$}
\affiliation{
$^{1}$Department of Material Science,
Graduate School of Material Science,\\
University of Hyogo\\
 Ako, Hyogo 678-1297, Japan\\
$^{2}$Faculty of Education, Kumamoto University, Kurokami 2-40-1, 
Kumamoto, 860-8555, Japan
}

\date{\today}


\begin{abstract}
In quasi-one-dimensional conductors, the conductivity in both one-dimensional axis and interchain direction
shows peaks when magnetic field is tilted at the magic angles in the plane perpendicular to the conducting chain.
Although there are several theoretical studies to explain the magic angle effect,
no satisfactory explanation, especially for the one-dimensional conductivity, has been obtained. 
We present a new theory of the magic angle effect in the one-dimensional conductivity
by taking account of the momentum-dependence of the Fermi velocity, 
which should be large in the systems close to a spin density wave 
 instability.
The magic angle effect is explained in the semiclassical equations of motion,
but neither the large corrugation of the Fermi surface due to long-range hoppings nor
hot spots, where the relaxation time is small, on the Fermi surface are required. 

\end{abstract}

\maketitle


Quasi-one-dimensional organic conductors    
show a lot of interesting phenomena, such as unconventional
superconductivity, spin density wave (SDW), field-induced spin density wave, quantum Hall effect,
etc\cite{IshiguroYamaji}. 
By studying the interaction of  electrons with impurities  
as well as interaction between electrons 
Lebed and Bak\cite{LebedBak1989} have predicted that the resistance, $R_{xx}$, 
has peaks at  the magic angles $\theta = \tan^{-1} (\frac{b}{c} \frac{p}{q})$,
where $p$ and $q$ are mutually prime integers and $b$ and $c$ are lattice constants, 
when the magnetic field is rotated in the plane perpendicular to
the most conducting $a$ axis.
Experimentally, dips, instead of peaks, of the resistance at the magic 
angles are observed 
in both $R_{xx}$ and $R_{zz}$ in various quasi-one-dimensional 
organic conductors,
(TMTSF)$_2$ClO$_4$\cite{Osada1991,Naughton1991}, 
(TMTSF)$_2$PF$_6$\cite{Kang1992,Chashechkina1998},  
(TMTSF)$_2$ReO$_4$\cite{Kang2003}, 
(DMET)$_2$I$_3$\cite{Uji1999}, 
(DMET)$_2$CuCl$_2$\cite{Ito2005}, (DMET-TSeF)$_2$AuCl$_2$\cite{Biskup2000} and 
(DMET-TSeF)$_2$AuI$_2$\cite{Biskup2000,Shimojo2002}. It is  called as the magic angle effect(MAE).

Other angle dependences of the magnoresistance have also been  observed in quasi-two-dimensional 
conductors\cite{Kartsovnik1988,Kajita1989,Iye1994}, and in quasi-one-dimensional conductors with rotating
magnetic field in the $a$-$c$ plane\cite{Danner1994} and in the $a$-$b$ plane\cite{Osada1996}.
These angle dependences are explained  in the semiclassical theory\cite{Yamaji1989,Yagi1990,Danner1994,Osada1996}
and they are used as  powerful tools to observe the shape of the Fermi surface
in low-dimensional systems.

For MAE, however,  no complete explanation  has  been given yet, 
although there exit several
theories\cite{Osada1992,Chaikin1992,Maki1992,Strong1994,Blundell1996,Moses2000,Lebed2003,Lebed2004,Lundin2004,Lebed2005}.  
Interesting idea of the
magnetic-field-induced confinement\cite{Strong1994} 
has been proposed to explain 
MAE. 
Difference between the ground states at the magic angles
and at other directions of the magnetic field is observed by the 
metallic and non-metallic temperature dependence of the resistance\cite{Chashechkina1998}.
In recent NMR experiment\cite{Wu2005}, however, no evidence for the 
change of the excitation spectrum at the magic angles has been obtained. This fact suggests that
semiclassical approach, or equivalently Green function approach using the linearized $k_x$ dispersion,
can be applied in these systems.
Indeed, MAE can be explained qualitatively in the
semiclassical theory.
By using the semiclassical equations of motion for 
the noninteracting electrons 
in the tilted magnetic field,
Osada et al.\cite{Osada1992} have shown that
the conductivity in the $z$ axis $\sigma_{zz}$
shows peaks at the magic angles.
In quasi-one-dimensional conductors the energy dispersion is 
described as
$\epsilon_{\bf{k}} = -2 t_a \cos{a k_x} - 2 t_b \cos b k_y - 2 t_c \cos c k_z - \mu.$
Here we take the simple cubic lattice with lattice constant $a$, $b$, 
and $c$ for simplicity.
Since $t_a \gg t_b > t_c$, the Fermi surface is two nearly parallel 
sheets at $k_x \approx \pm k_F$
and $k_F = \frac{\pi}{4 a}$ for the quarter-filled band. 
Osada et al\cite{Osada1992} used the linearized dispersion in the $k_x$ direction as
\begin{equation}
 \epsilon_L(\mathbf{k}) = \pm \hbar v_F (|k_x|-k_F) - \sum_{m,n} t_{mn} \cos (m b k_y+n c k_z),
\end{equation}
and constant relaxation time $\tau$.
They have shown that peaks appear in $\sigma_{yy}$ and $\sigma_{zz}$ at 
the magic angles   
and the heights of peaks are proportional to $t_{pq}^2$.
In order to explain the experimental 
results of the perpendicular magnetoresistance,
unphysically large values of $t_{pq}$ are required.
This model  also fails to explain the magnitude of the recently observed
 giant Nernst effect\cite{Wu2003}  at the magic angles.  
Maki\cite{Maki1992} has used 
the similar approximation of the linearized dispersion and 
constant $\tau$, 
but took account of the $k_y$ dependence of $v_x$ 
in perturbation in $\frac{t_b}{t_a}$.
Recently, Lebed et al.\cite{Lebed2004,Lebed2005} 
have emphasized the importance of the $k_y$ dependence of $v_x$, or the density of states on the Fermi surface.
 They show that there occurs the 1D to 2D 
dimensional crossover when  the magnetic field is at magic angles. 
In these approximation resonance-like peaks 
in $\sigma_{zz}$ at the magic angle with $q=1$ is obtained, but  its height is proportional to
$(\frac{t_b}{t_a})^{2p}$ and very small in the case of $t_b \ll t_a$\cite{Maki1992}.

The more serious discrepancy between the semiclassical theories of MAE and experiments is 
encountered in the
one-dimensional conductivity $\sigma_{xx}$. 
In the model of Osada et al.\cite{Osada1992},  $\sigma_{xx}$ 
does not depend on the direction
of the magnetic field. 
Maki\cite{Maki1992} has pointed out that $\sigma_{xx}$  shows peaks at 
magic angles,
if the $k_y$ and $k_z$ dependence of $v_x$ is taken into account.
 It is, however,  very weak, i.e., it is of the order of $(\frac{t_c}{t_a})^2$. 
%
Meanwhile,
Chaikin\cite{Chaikin1992} has shown that the angular-dependence 
of $\sigma_{xx}$ can be
explained, if there are \textit{hot spots} on the Fermi surface, where
 $\tau$ is small.

In this letter we show  that the peaks in  one-dimensional conductivity
 $\sigma_{xx}$ at the magic angles are dramatically enhanced, when
$v_x$ depends strongly on $k_y$ and $k_z$ 
 on the Fermi surface, even when long-range hopping terms and hot spots are absent.
In quasi-one dimensional organic conductors, 
 electrons correlate strongly and the system is close to the
insulating state of SDW or field-induced SDW
due to the nearly perfect nested Fermi surface. 
Then  $v_x$ on the Fermi surface depends  on $k_y$ and $k_z$ 
much stronger than  in perturbation in $\frac{t_b}{t_a}$ and $\frac{t_c}{t_a}$ 
studied by Maki\cite{Maki1992} and Lebed et al.\cite{Lebed2003,Lebed2004}.
The effect of electron correlation on $v_x$ and $\tau$ can be calculated 
  by using the random phase approximation for the
quasi-one dimensional Hubbard model, which we will show in another publication\cite{kaneyasu2005}. 
 In order to show MAE due to the $k_y$ and $k_z$ dependence of $v_x$, we calculate $\sigma_{xx}$ in the model
 that
 the  velocity $v_x$ in some region on the Fermi surface differs sufficiently from that 
in the other region.

We assume the constant $\tau$. Then
the conductivity is calculated as\cite{Osada1992,Blundell1996,Osada1996,Moses2000}
\begin{equation}
\sigma_{ij}=\frac{2e^2}{V}\sum_{\mathbf{k}}\left(- \frac{df}{d \epsilon_{\mathbf{k}}} \right) v_i({\mathbf{k}}(0))
\int_{-\infty}^{0} v_j({\bf k}(t))e^{\frac{t}{\tau}} dt
\label{sigmaij}
\end{equation}
where $i$ and $j$ are $x$, $y$ or $z$, 
and $v_i({\bf k}(t))$ is the velocity on the Fermi surface in the semiclassical
picture
\begin{equation}
\mathbf{v}(\mathbf{k}(t))
 = \frac{1}{\hbar} \frac{\partial E(\mathbf{k})}{\partial \mathbf{k}(t)}.
\label{eqofv}
\end{equation}
The velocity $\mathbf{v}(\mathbf{k}(t))$ depends on time via the
time dependence of the wave vector $\mathbf{k}(t)$ as
\begin{equation}
 \hbar \frac{d \mathbf{k}(t)}{d t} = -e \mathbf{v}(\mathbf{k}(t)) \times \mathbf{B}
\label{eqofk}
\end{equation}
where we take the magnetic field $\mathbf{B}$  in the $y-z$ plane,
$ B_x = 0$,
$ B_y = B \sin \theta$, and
$ B_z = B \cos \theta$.

\begin{figure}[tb]
\includegraphics[width=0.4\textwidth]{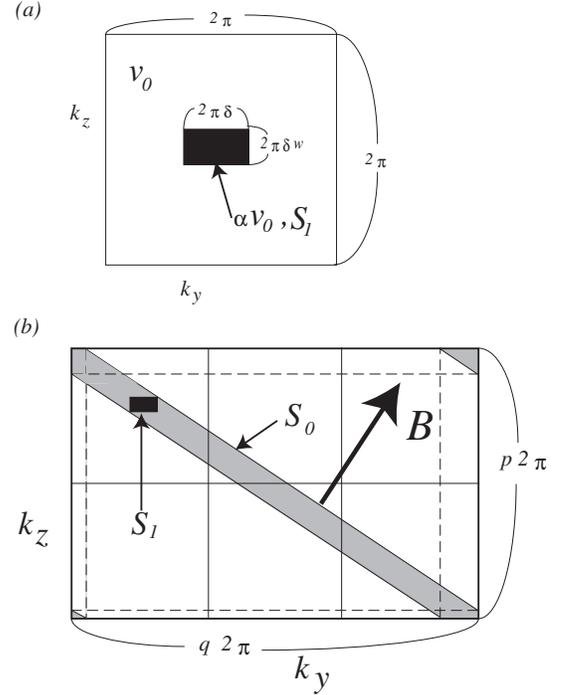}
\caption{(a) Simple model that $v_x$ is $\alpha v_0 $ in the
small rectangular region and $v_0$ in the other region in the
Brillouin zone.
(b) Extended Brillouin zone for $\tan \theta = \frac{p}{q} = \frac{2}{3}$.
The gray region should be translated to the first Brillouin zone. 
}
\label{fig1}
\end{figure}

When $\theta = \tan^{-1} (\frac{b}{c} \frac{p}{q})$, $\mathbf{k}$ started from some point $\mathbf{k}(0)$
 moves in the commensurate trail on the Fermi surface before coming back to $\mathbf{k}(0)$,
while it travels all the Fermi surface when $\frac{c}{b}\tan \theta$ is an irrational number.
This is the main idea of the Chaikin's hot spots and other semiclassical theories of MAE.
In order to simplify the calculation, we assume $v_x=v_0$ except for in the rectangular region of $2 \pi \delta 
\times 2 \pi \delta w$, where $v_x=\alpha v_0$ ($\alpha \neq 1$), as shown in Fig.~\ref{fig1}(a)
and $a = b = c = 1$.
%
%
In the quasi-one dimensional system the density of states 
on the Fermi surface can be expressed as a function
of energy $\epsilon$,  $k_y$ and $k_z$ as
\begin{align}
 N(\epsilon, k_y, k_z) 
 &= \frac{1}{\left| \hbar v_x(\epsilon, k_y, k_z)\right|}
\end{align}
At low temperature we can replace $- \frac{\partial f}{\partial \epsilon_{\mathbf{k}}}$ by
$\delta(\epsilon - \epsilon_F(k_y, k_z))$ in Eq.(\ref{sigmaij}) and we obtain
\begin{align}
 \sigma_{xx} 
&= 
2 \frac{2 e^2}{\hbar V}  \sum_{k_y, k_z} \int_{-\infty}^{0} v_x(\mathbf{k}(t)) e^{\frac{t}{\tau}} dt,
\end{align}
where summation on $k_y$ and $k_z$ should be done on the right 
sheet of the Fermi surface ($v_x(\mathbf{k}(t)))>0$).
A factor of $2$ comes from the contribution of the left sheet of the Fermi surface.
Although $\sigma_{zz}$ is  given by the correlation of $v_z(\mathbf{k}(0))$ and $v_z(\mathbf{k_z}(t))$
in  Eq.(\ref{sigmaij}), $\sigma_{xx}$ in quasi-one dimensional system is simply given by the
momentum-space average of the
time average (weighted by $\exp(\frac{t}{\tau})$) of $v_x(\mathbf{k}(t))$.  
We consider the case of $eB\tau \gg 1$. Then integration over $t$ can be approximated by
$\tau$ times the 
time average of $v_{x}(\mathbf{k}(t))$ over the trajectory of $\mathbf{k}(t)$.
It might be thought that $\sigma_{xx}$ does not depend on the direction of the magnetic field,
but it does, since the time-average is performed on the trajectory,
which depends on the angle of the magnetic field.
%
%
Since the speed of $\mathbf{k}(t)$  
in the momentum space is proportional to $v_x(\mathbf{k}(t))$ 
when $|v_x| \gg |v_y|, |v_z|$, as seen from Eq.~(\ref{eqofk}),
time average of  $v_x(\mathbf{k}(t))$ can be calculated as follows.

First,
we consider the commensurate case,
$ \tan \theta = \frac{p}{q}$,
where $p$ and $q$ are mutually prime integers.
Brillouin zone is extended $q$ times $p$ as shown in Fig.~1(b).
In the black region (area $S_1=(2 \pi)^2 w \delta^2 $), $v_x= \alpha v_0 $. 
The gray and black regions in Fig.~1(b) is defined as the set of $\mathbf{k}(0)$ that   goes into the
black region 
at  some $t$.
The area of the gray region  is 
\begin{eqnarray}
S_0 &=& (2 \pi )^2  p (1+\frac{w}{\tan \theta} ) \delta - S_1 \nonumber \\
  &=& (2 \pi)^2  (p+q w - \delta w) \delta .
\end{eqnarray}
We define 
$s_0 \equiv \frac{S_0}{(2\pi)^2}= (p+q w - \delta w) \delta $
 and 
$s_1 \equiv \frac{S_1}{(2\pi)^2} =   w \delta^2 $.
The gray region in Fig.~\ref{fig1}(b)  should be translated in the first Brillouin zone.
Therefore, if $(p+q w) \delta >1$, all the first Brillouin zone becomes black or gray region, i.e.,
every $\mathbf{k}(0)$ goes into $S_1$. If   $ (p+q w) \delta < 1$, however, we can divide the first Brillouin zone 
into two regions,  the first region ($S_1+S_0$), and the second region $(2\pi)^2-(S_1+S_1)$.
If $\mathbf{k}(0)$ is in the first region, $v_x(\mathbf{k}(t))$ changes the value as $\mathbf{k}(t)$ 
travels on the commensurate trajectory on the Fermi surface, 
while $v_x(\mathbf{k}(t))$ is time-independent constant
$v_0$ for the $\mathbf{k}(0)$ in the second region.
Now, we calculate the momentum-space average and time average of 
$v_x$, when the magnetic field is at the magic angle 
$\theta= \tan^{-1}\frac{p}{q}$ and $ (p+q w) \delta<1$. 
If $\mathbf{k}(0)$ is in $S_1$ or  $S_0$,  the duration when $\mathbf{k}(t)$ is in the region 
$S_1$  is approximately  proportional to $\frac{s_1}{\alpha v_0}$, 
while that in $S_0$ is approximately
proportional to $\frac{s_0}{v_0}$.
Thus the time average of $v_x(\mathbf{k}(t))$ when $\mathbf{k}(0)$ is in $S_1+S_0$ is given by
\begin{align}
\langle v_x \rangle^{(S_1+S_0)} &\approx
 \alpha v_0  \frac{\frac{s_1}{\alpha v_0}}{\frac{s_1}{\alpha v_0}+ \frac{s_0}{v_0}}
 + v_0      \frac{\frac{s_0} {v_0}}     {\frac{s_1}{\alpha v_0}+ \frac{s_0}{v_0}}
\nonumber \\ 
&= v_0 \frac{s_1+s_0}{\frac{s_1}{\alpha}+s_0}
\end{align} 
After the summation over $\mathbf{k}(0)$ in the Brillouin zone,
 we get the time and momentum-space average of $v_x$ for the magic angle (commensurate orbit) as
\begin{align}
 \langle v_x \rangle^{C} &=
v_0 (1-s_1 - s_0)+ \langle v_x \rangle^{(S_1+S_0)} (s_1+s_0)
\nonumber \\
 &= v_0 \left( 1 + \frac{s_1 (s_1+s_0) (1-\frac{1}{\alpha})}{\frac{s_1}{\alpha} + s_0}
 \right)
\end{align}
%
\begin{figure}[tb]
\includegraphics[width=0.4\textwidth]{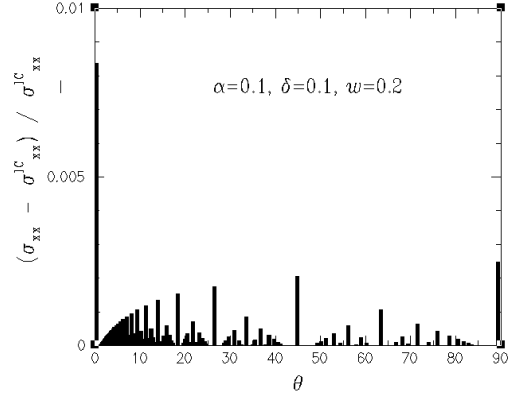}
\caption{Angular dependence of $\sigma_{xx}$ for the model of  $k_y$ and $k_z$ dependent $v_x$. }
\label{sxx010102}
\end{figure}
\begin{figure}[tb]
\includegraphics[width=0.4\textwidth]{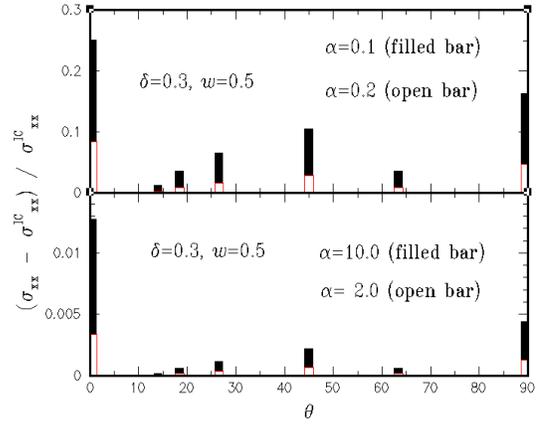}
\caption{Angular dependence of $\sigma_{xx}$ for another choice of $\alpha$, $\delta$ and $w$. 
In this case fewer peaks are seen but peak heights are large.}
\label{fig34}
\end{figure}
%
Next, the time and momentum-space average of $v_x$ is calculated for the case 
that $\tan \theta$ is irrational number or $( p+ q w) \delta >1$. In this  case,
$\mathbf{k}(t)$ goes into the region $S_1$ at some $t$ regardless of the initial point 
$\mathbf{k}(0)$, i.e. the trajectory is incommensurate. The average of $v_x$ is calculated as
\begin{align}
 \langle v_x \rangle^{IC} 
&= \alpha v_0 \frac{\frac{s_1}{\alpha v_0}}{\frac{s_1}{\alpha v_0} +\frac{1-s_1}{v_0}} + 
  v_0 \frac{\frac{1-s_1}{v_0}}{\frac{s_1}{\alpha v_0} +\frac{1-s_1}{v_0}}
\nonumber \\
&= v_0 \frac {1}{1-s_1(1-\frac{1}{\alpha})}
\end{align} 

From the above result we get the normalized peak heights of conductivity at the magic angles as 
\begin{align}
\frac{\sigma_{xx}-\sigma_{xx}^{IC}}{\sigma_{xx}^{IC}} &=
\frac{\langle v_x \rangle^C - \langle v_x \rangle^{IC}}{\langle v_x \rangle^{IC}}
\nonumber \\
&= \frac{(1-\alpha)^2 w^2 \delta^3}{\alpha} \frac{1-(p+q w)\delta }{\alpha(p+q w) + w (1-\alpha)\delta},
\end{align}
where the condition $(p+q w) \delta <1 $ should be satisfied.

In both cases of small velocity ($\alpha <1$) and large velocity ($\alpha >0$)
in region $S_1$, we get $\sigma_{xx} \geq \sigma_{xx}^{IC}$.
The peak heights becomes large ($\propto \frac{1}{\alpha}$) as $\alpha$ approaches to $0$.  
In Figs. \ref{sxx010102} and \ref{fig34}, we plot $(\sigma_{xx} -\sigma_{xx}^{IC})/\sigma_{xx}^{IC}$
for several parameters of $\alpha$, $\delta$ and $w$.
If $s_1$ is small, there are a lot of peaks
at magic angles,
$\tan^{-1}1=45^{\circ}$, 
$\tan^{-1}2\approx 63.4^{\circ}$, 
$\tan^{-1}\frac{1}{2}\approx 26.6^{\circ}$,
$\tan^{-1}\frac{1}{3}\approx 18.4^{\circ}$,
etc.
as seen in Fig.~\ref{sxx010102}, but the heights of peaks are small.
On the other hand, the number of peaks is small but their heights are large for larger $s_1$ 
(Fig.~\ref{fig34}).

In conclusion we show that peaks in $\sigma_{xx}$ at magic angles are large,
 when $v_x$ depends on $k_y$ and $k_z$
on the Fermi surface of the quasi-one dimensional conductors. In the simple model that $v_x=v_0$ is constant except
for in the rectangular region on the Fermi surface, where $v_x$ is another constant value $v_x=\alpha v_0$,
we get the approximate form of $\sigma_{xx}$ by using the simple calculation of taking the 
momentum-space average 
on the Fermi surface and the time average on the trajectory of $\mathbf{k}(t)$.
MAE is the consequence of the fact that $\mathbf{k}(t)$ stays shorter time
in the region where $|v_x(k_y, k_z)|$ is large,  and in the region of
small $|v_x(k_y,k_z)|$,  $\mathbf{k}(t)$ stays longer time. 
In this Letter, only $\sigma_{xx}$ as a function of the tilting angle of the magnetic field is studied. 
The similar mechanism works also for  $\sigma_{zz}$ and 
the peak heights of $\sigma_{zz}$  at the  magic angles will be much more enhanced 
by considering the strong $k_y$ and $k_z$  dependence of $v_x$  as done in this paper 
than that studied by Maki\cite{Maki1992} and 
Lebed et al.\cite{Lebed2004,Lebed2005} in the perturbation theory in $\frac{t_b}{t_a}$. 
If we take account of the momentum-dependent relaxation time and more realistic
momentum dependence of the velocity $v_x(k_y, k_z)$, experimentally observed peaks of $\sigma_{xx}$ 
and $\sigma_{zz}$
at magic angles will be explained better.
Although MAE of $\sigma_{xx}$ can be explained within the semiclassical theory, 
 the strong momentum-dependence
of the Fermi velocity, which may be caused by 
 interactions between electrons, plays a crucial role to understand the magnitudes of the peak heights.
This feature is in contrast to the
other  angle dependence of the conductivity in quasi-one and quasi-two dimensional
conductors\cite{Yamaji1989,Yagi1990,Danner1994,Osada1996} 
and usual quantum oscillations such as Shubnikov-de Haas
oscillations, where the shape of the Fermi surface is important.

This work is partly supported by a Grant-in-Aid 
for the Promotion of Science and Scientific Research on
Priority Areas (Grant No. 16038223) from the Ministry of
Education, Culture, Sports, Science and Technology.
%

\end{document}